\documentclass{article}
\usepackage{amsmath}
\usepackage{graphicx}
\usepackage{cite} 
\usepackage{authblk}
\usepackage[a4paper, total={15.0cm, 26.0cm}]{geometry}

\title{Orbital and spin current density backflow in unidirectional 
monochromatic electromagnetic fields in vacuum}

\author[1,2,*]{Peeter Saari} 
\author[3]{Ioannis Besieris}

\affil[1]{Institute of Physics, University of Tartu, Tartu 50411, Estonia}

\affil[2]{Estonian Academy of Sciences, Tallinn 10130, Estonia}

\affil[3]{The Bradley Department of Electrical and Computer Engineering, 
Virginia Polytechnic Institute and State University,
Blacksburg, Virginia 24060, USA.}

\affil[*] {peeter.saari@ut.ee}

\begin{document}
\maketitle
\begin{abstract}
In this study, energy backflow in the Poynting vector, as well as its orbital and spin current 
density components, has been examined for a 2-dimensional causal unidirectional 
vector-valued monochromatic electromagnetic wave. Linear transverse electric (TE), transverse 
magnetic (TM), and circular polarization cases are considered and studied in detail, including 
both electric and magnetic contributions to the current density components. Spin current backflow 
has been found to be unexpectedly strong.  A study of the energy backflow is also presented in 
the scalar version of the 2-dimensional monochromatic wave. A detailed study has been carried out 
of the correlation of the positions of energy backflows with local wavenumbers and their signs, 
the zeros of appropriate intensities and the presence of vortices.
\end{abstract}

\section{Introduction}
The phenomenon of energy backflow or reverse (negative) flow takes place when the Poynting 
vector is directed backward with respect to its predominant direction. If the wavefield propagates 
in the direction of the $z$-axis, it means that the $z$-component of the Poynting vector is negative
($P_z<0$) at some spatio-temporal regions. 

Energy backflow in tightly focused light beams has been known for a long time 
\cite{ignatovsky1919diffraction,richards1959electromagnetic}, but has become the subject of new research activity (see \cite{kotlyar2019energy,kotlyar2020mechanism,li2020controlled,han2024controllable} and references
therein).

At the end of the last century it was shown that backflow appears in a superposition of as few as 
four appropriately polarized and directed electromagnetic plane waves \cite{katsenelenbaum1997direction}. 
Very recent thorough study of such a quartet of waves points out the crucial role of non-paraxiality and 
polarization in the appearance of the backflow \cite{saari2021backward,Chun-Fang2026}. The role of these
characteristics appears also in the formation of the backflow in Bessel beams \cite{novitsky2007negative,
mitri2016reverse} and their pulsed superpositions \cite{salem2011energy}. The list of waves where the 
phenomenon is possible has expanded and incorporates Airy beams \cite{vaveliuk2012negative}, Lissajous 
beams\cite{khonina2022tailoring}, and various localized space-time wave packets \cite{besieris2023energy}.

In \cite{bialynicki2022backflow} proof is given that the backflow is a wave phenomenon that may occur 
in all kinds of wavefields describable by different types of wave equations. In particular, the
example of the  hopfion solution of Maxwell equations in the appendix of \cite{bialynicki2022backflow} reveals 
the presence of energy backflow in the case of unidirectional waves. Indeed, the phenomenon is 
especially counterintuitive if the Poynting vector is directed backward with respect to the 
directions of all plane-wave constituents of the wavefield. For this reason, in our recent studies 
\cite{besieris2023energy,saari2024energy} we focused on the blackflow in unidirectional waves.

The sharp increase in the number of publications on optical backflow in recent years is caused 
by the relevance of the phenomenon to topics such  as particle manipulation, superoscillations, and 
superresolution (see, e.g.\cite{geints2022simulation,yuan2019plasmonics} and references therein). 
The close relation of the backflow to optical vortices has been intensively studied in the last 
couple of years
\cite{kotlyar2025canonical,kotlyar2026reverse,kotlyar20262d,kotlyar2026reverseAPB,kotlyar2025backward,stafeev2026phase}

The Poynting vector can be decomposed into the sum of two vectors, describing respectively 
the orbital energy flux (canonical energy flow) and the spin flow 
\cite{bekshaev2007transverse,berry2009optical,ghosh2024canonical,ustinov2024interference}. In order 
for such a splitting to be unique, it requires "electric–magnetic democracy," 
which means expressing in turn  both parts as sums of electric and magnetic terms \cite{berry2009optical,
bliokh2014extraordinary}.

In \cite{kotlyar2026reverse,kotlyar20262d} reverse energy flows in  non-paraxial  monochromatic 
2-dimensional (2D) transverse electric (TE) fields have been studied by Kotlyar \textit{et al}. 
The fields they studied---transverse Bessel and sinc-beam---are unidirectional but 
the spin flow component of the Poynting vector is absent due to TE polarization and neglecting the
magnetic field contribution to spin flow. The canonical (orbital) component of the Poynting 
vector has been calculated in these papers also without taking into account "electric–magnetic democracy." 
A remarkable result obtained by Kotlyar \textit{et al.} is that the 2D fields exhibit energy backflow 
near the intensity zeros of the field and energy flow vortices.

Our aim is to study in detail both orbital and spin backflows, and separately contributions to them
according to a "EM-democratic" approach in the case of the  monochromatic 2D unidirectional wavefunction 
we found in \cite{saari2024energy}. Our wave function is an exact closed-form expression over the whole
2D space, in contrast to Refs. \cite{kotlyar2026reverse,kotlyar20262d} . This provides us great facility in computing, without resorting to numerical integrations, the Poynting vector, 
as well as its orbital and spin current density components, and graphing the results for any relevant 
value of $z$. 

This article is organized as follows. For the reader's convenience, in section 2 we briefly describe 
the two-dimensional unidirectional wave function derived in \cite{saari2024energy}. Then, we start by 
studying the backflow of a scalar wave current, not because it is simpler from the energy backflows 
of the vector-valued fields that will be studied later on, but because we wish to start a systematic 
pattern of presentation of graphical results of the backflow itself, the location of backflow peaks,
corresponding flow vortices, and “superoscillations”.

In section 4 we start with a summary of the decomposition of the Poynting vector into its
orbital and spin components. Then follow two subsections devoted to the main results of the paper---
study of electric and magnetic parts and their joint effect on the backflow in transverse electric 
(TE), transverse magnetic (TM), and circularly polarized fields. In section 5, we summarize the 
results and present concluding remarks.

\section{Two-dimensional monochromatic unidirectional wave}
In our recent paper \cite{saari2024energy} we succeeded in finding a closed-form 
expression with fractional order Bessel functions 
\begin{equation}
W(x,z)=C\sqrt{k\,\left\vert z\right\vert }\left\{
\begin{array}
[c]{c}%
J_{-\frac{1}{4}}\left[  \frac{k}{2}\left(  \sqrt{x^{2}+z^{2}}-x\right)
\right]  \,\\
\times J_{-\frac{1}{4}}\left[  \frac{k}{2}\left(  \sqrt{x^{2}+z^{2}}+x\right)
\right]  \\
+i\,signum\left(  z\right)  \,\\
\times J_{\frac{1}{4}}\left[  \frac{k}{2}\left(  \sqrt{x^{2}+z^{2}}-x\right)
\right]  \,\\
\times J_{\frac{1}{4}}\left[  \frac{k}{2}\left(  \sqrt{x^{2}+z^{2}}+x\right)
\right]
\end{array}
\right\}  , \label{Wanalyt}%
\end{equation}
which, upon multiplication with $e^{(-ikct)}$ describes a unidirectional wave propagating
along the $z$-direction. 
Here $k$ is the wavenumber, taken equal to $\pi$ throughout this paper which gives to the 
wavelength a value $\lambda=2\pi/k=2$ dimensionless units, correspondingly allowing coordinate 
scales of the figures below to be converted to real optical wavelength in angstroms, nanometers, 
etc., used in practice. The constant $C=\Gamma(3/4)^{2}/2\approx0.75$, 
where $\Gamma$ is the gamma function, ensures normalization of the field to unity at the origin. 
Let us  notice that since the arguments of the Bessel functions are non-negative, they have no 
imaginary parts, and a simple complex conjugation symmetry holds \mbox{$W(x,z<0)=W^{\ast}(x,z>0)$}.
The angular spectrum used in the Fourier synthesis of $W(x,z)$, is $S(\theta)=1/\sqrt{\cos\theta}$ 
if $0<\theta<\pi/2$ and $S(\theta)=0$ otherwise, where $\theta$ is the angle between the wave 
vector of a  constituent plane-wave and the axis $z$. Hence, the wave is unidirectional and highly 
non-paraxial with increasing weight of plane waves incident at large angles to the $z$-axis. 
The unidirectionality of $W(x,z)$ was additionally proved by discrete Fourier analysis of its 
behavior along the propagation axis. The modulus squared of $W(x,z)$
consists of a peak at the origin and exhibits low-intensity oscillations in the $(x,z)$-plane (see a 
3D plot in \cite{saari2024energy}).
The wavefunction of $W(x,z)$ from Eq.~(\ref{Wanalyt}) is the basis for all calculations in this study.
\section{Scalar wave current backflow}
The energy current in a time-dependent scalar field $\Psi$ is generally given by expression \cite{mandel1996optical}:
\begin{equation}
   \overrightarrow{J}  =-\frac{1}{2}\left(  \partial_{ct}\Psi^{\ast}\right)  \left(
\nabla\Psi\right)  -\frac{1}{2}\left(  \partial_{ct}\Psi\right)  \left(
\nabla\Psi^{\ast}\right)~,\label{Jgen}
\end{equation}
where $\Psi^*$ is the complex conjugate of $\Psi$.
In the case of a monochromatic wave with wave vector $k$, $\Psi(x, y,z,t)$ can be expressed in terms of its modulus, spatial phase, and time exponent, which in our 2D case reads
\begin{equation}
\Psi(x, z) = |W(x, z)|\, \exp[i \chi(x, z)] \,\exp(-ikct). 
\end{equation} \label{mod2exp}
From Eq~(\ref{mod2exp}) it follows \cite{berry2009optical} that
\begin{equation}
\overrightarrow{J}(x, z)/k = |W(x, z)|^2 \;\nabla \chi(x, z) \hspace{0.7 cm} \mbox{or} \hspace{0.7 cm} 
\overrightarrow{J}(x, z)/k = \text{Im}\{W^*(x, z)\, \nabla W(x, z)\}, 
\end{equation}\label{J2x}
which can be proved by a simple vector calculus derivation.
The wave number $k$ can be omitted because, depending on the units used, the definition of current 
usually contains a constant factor. The curl of $\overrightarrow{J}$ is referred to as the vorticity 
vector \cite{berry2009optical}.

Figure \ref{Jz1} below illustrates the positive and negative (backflow) values of the longitudinal 
projection of the current vector. We see that by comparing the maxima, the backflow is 25 times weaker
than the forward flow.
\begin{figure}[h]
\centering
\includegraphics[width=14cm]{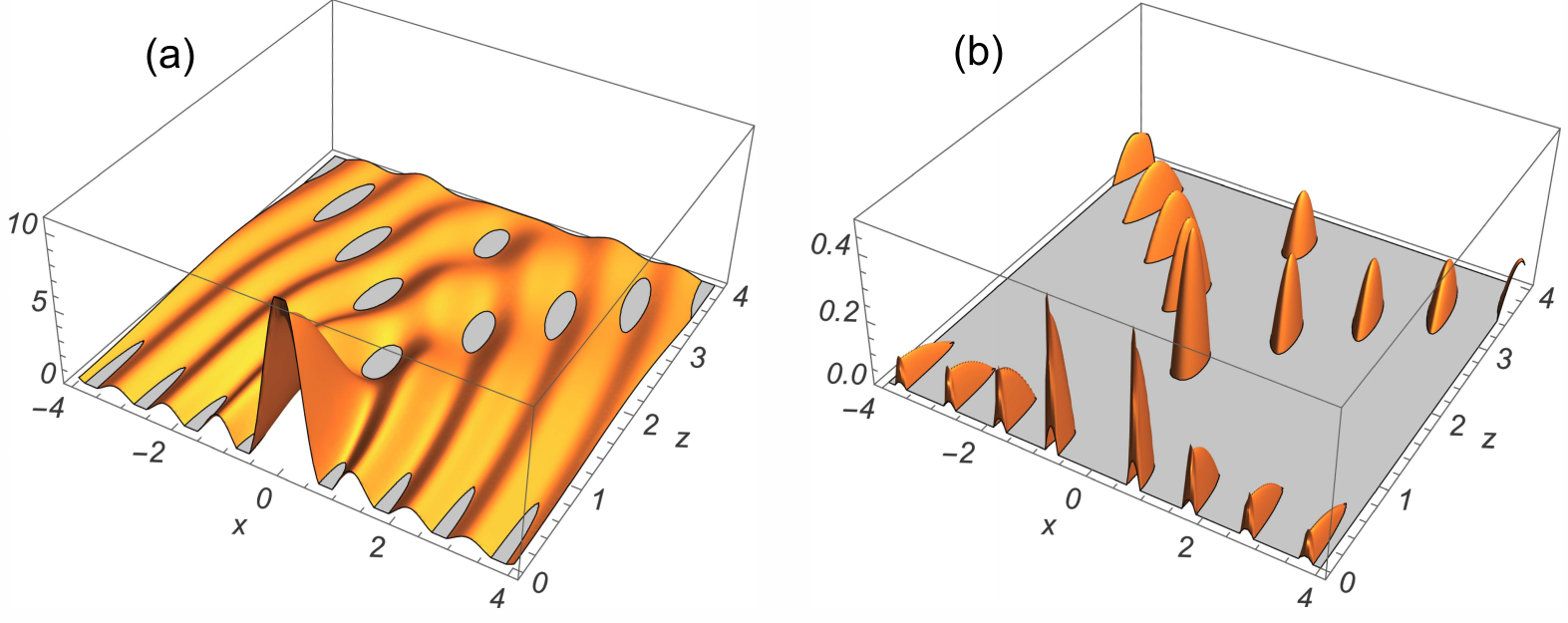}\caption{Plot of the longitudinal ($z$) projection of the 
current vector $\overrightarrow{J}$, see Eq. (\ref{J2x}); (a), only 
positive values of the projection are shown, (b), only negative values, i.e., the backflow, are shown. 
The unit of spatial coordinate scales is $\lambda/2$. The vertical scale is in relative units.} 
\label{Jz1}
\end{figure}
\begin{figure}[h!]
\centering
\includegraphics[width=14cm]{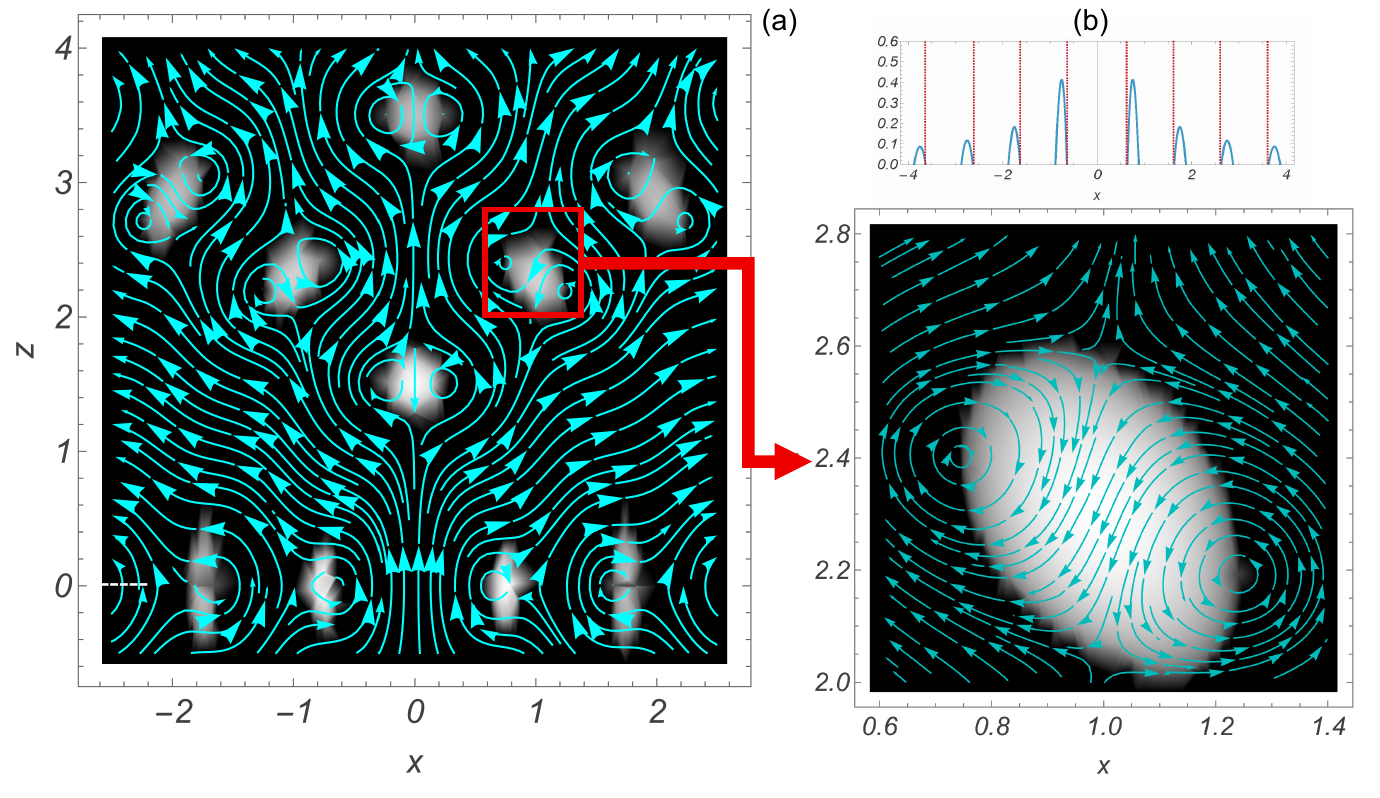}\caption{(a), greyscale density plot of negative values of the 
longitudinal  $z$-projection (backflow) $(-J_z)^{\frac{1}{3}}$ (the quantity has been raised to the 
power 1/3 in order to make otherwize too dark peaks brighter), superimposed by a stream plot of the 
vector field $(J_x,J_z)$. Unlike figure \ref{Jz1}b, the axis $z$ starts from $z=-0.5$ to make the backflow
peaks located along the line $z=0$ fully visible. On the right an inset of a small area 
$0.8\lambda\times0.8\lambda$ containing one backflow peak is shown with higher resolution. (b), 
backflow peaks (solid curve) along the line $z=0$, see figure \ref{Jz1}b, juxtaposed with locations of zeros
(dashed verticals) of $W(x, z)$ along the same line $z=0$.}\label{strJ}
\end{figure}
Figure \ref{strJ} shows that backflow peaks are accompanied by vortices and---more exactly---they are 
located between vortices of opposite topological charge. The peaks along the line $z=0$, on the contrary,
are accompanied with vortices only from one side. The reason for this is clear from the figure 
\ref{strJ}b, where zeros of 
\begin{equation}
W(x,z=0)\propto\left[3 J_{3/4}(kx)-2kxJ_{7/4}(kx)\right]/(kx)^{3/4} \label{z=0}
\end{equation}
are shown (see Eq.~(11) 
from Ref.~\cite{saari2024energy}). Namely, as it is well known and as we see also below, the vortices are
located at zero points of intensity, which along the line $z=0$ are located only on one side of the 
backflow peaks.

A stream plot on the same plane $(x,z)$ 
of the vector field $(J_y,J_z)$ has no vortices and consists of arrows all parallel to the $z$-axis
because $J_y \equiv0$. But the arrows flip their direction along the boundary line between 
the forward flow and backflow areas revealing the sign of $J_z$.

\begin{figure}[h!]
\centering
\includegraphics[width=14cm]{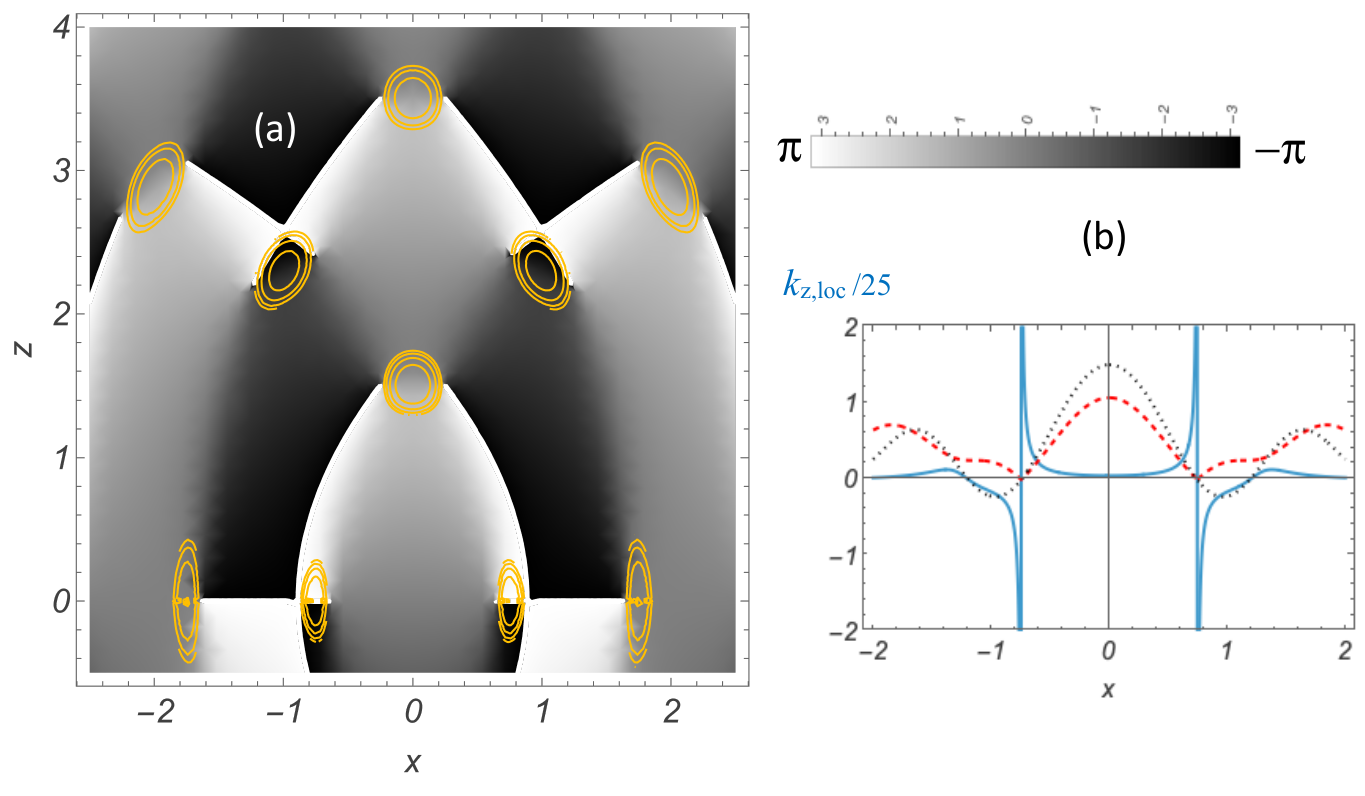} \caption{(a),  contour plot of negative values of the longitudinal
$z$-projection (backflow) $(-J_z)^{\frac{1}{3}}$, superimposed by greyscale plot of the phase function
$\chi(x, z)$ from Eq.~(\ref{mod2exp}); (b), dependence of $k_{z,loc} \equiv \partial\chi(x, z)/\partial z$
along the line $z=2.4$ (solid curve) in comparison with the modulus $|W(x, z)|$ of the wave function
(dashed curve), and $J_z$ (dotted curve), both along the same line.} \label{gigaJ}
\end{figure}

Figure \ref{gigaJ}a shows that the backflow maxima are located in close proximity to abrupt phase changes.
Figure \ref{gigaJ}b in turn demonstrates that phase jumps take place at intensity zeros---it is generally
known that physically they cannot be located elsewhere. When juxtaposing figure \ref{gigaJ}b and the inset
of figure \ref{strJ}, we observe that the center of the vortex is located precisely at the point 
where both the intensity is zero and the phase derivative exhibits a singularity. Naturally, this point 
is also crossed by the boundary line between the forward flow and backflow areas. The coordinates of 
all four features are (0.75, 2.4) and (-0.75, 2.4). According to Eq.~(\ref{J2x}), $k_{z,loc} \equiv 
\partial\chi(x, z)/\partial z$ and $J_z$ must have the same sign as confirmed by figure \ref{gigaJ}b.

The reason for phase jumps far from the backflow 
maxima, e.g., along the line $z=0$ between two backflow maxima, is technical: both the computer functions
Arg($Z$) and Log($Z/\vert Z\vert)/i$ return the phase of a complex number $Z$ modulo $2\pi$, i.e., 
as soon as the phase increases over $\pi$ it jumps to $-\pi$.

In general, although the current vector $\overrightarrow{J}$  and $\nabla\chi$ have the same streamlines,
the former is a smooth function at the centers of the vortices, whereas the latter diverges there. 
In figure \ref{gigaJ}b the absolute value of the local wavenumber $k_{z,loc} 
$ exceeds the value $10^2$, i.e, it is much larger than $k=\pi$. This is the 
phenomenon of the wave number becoming gigantic at intensity zeros (see \cite{kotlyar20262d} and
refrences therein). As the phenomenon takes place within sub-wavelength regions,  optical
“superoscillations” occur there (see \cite{berry2009optical} and references therein). These 
observations highlight spectacular phenomena in the field of singular optics.

\section{Orbital and spin current backflow}

\subsection{Splitting the Poynting vector into orbital and spin contributions and their electric and magnetic 
parts}

Consider the real time-harmonic electric and magnetic fields%

\begin{equation}
\overrightarrow{e}\left(  \overrightarrow{r},t\right)  =\operatorname{Re}%
\left\{  \overrightarrow{E}\left(  \overrightarrow{r}\right)  \,e^{-i\omega
t}\right\}  ,\quad\overrightarrow{h}\left(  \overrightarrow{r},t\right)
=\operatorname{Re}\left\{  \overrightarrow{H}\left(  \overrightarrow{r}%
\right)  \,e^{-i\omega t}\right\}  ,\label{1}
\end{equation}
respectively, in free space. The time-averaged Poynting vector is given by%
\begin{equation}
\overrightarrow{P}\left(  \overrightarrow{r}\right)  =\left\langle
\overrightarrow{e}\left(  \overrightarrow{r},t\right)  \times
\overrightarrow{h}\left(  \overrightarrow{r},t\right)  \right\rangle
_{time~aver.}=\frac{1}{2}\operatorname{Re}\left\{  \overrightarrow{E}%
^{\ast}\left(  \overrightarrow{r}\right)  \times\overrightarrow{H}\left(
\overrightarrow{r}\right)  \right\}  \label{2}
\end{equation}
This expression can be rewritten in the following two different forms, the
first involving only the electric field and the second only the magnetic
field:

\begin{equation}
\overrightarrow{P}\left(  \overrightarrow{r}\right)  =\varepsilon_{0}%
\frac{c^{2}}{2\omega}\operatorname{Im}\left\{  \overrightarrow{E}^{\ast}%
\times\left(  \nabla\times\overrightarrow{E}\right)  \right\}  =\mu_{0}%
\frac{c^{2}}{2\omega}\operatorname{Im}\left\{  \overrightarrow{H}^{\ast}%
\times\left(  \nabla\times\overrightarrow{H}\right)  \right\}  ,\label{3}
\end{equation}
where $c=1/\sqrt{\varepsilon_{0}\mu_{0}}$.

Below, a decomposition will be used of the time-averaged
Poynting into orbital and spin  current densities. In terms of only the
electric field one has%
\begin{align} \label{4}
\begin{split}
\overrightarrow{P}\left(  \overrightarrow{r}\right)    & =\overrightarrow{P}%
_{oe}+\overrightarrow{P}_{se}\,;\\
\overrightarrow{P}_{oe}  & =\varepsilon_{0}\frac{c^{2}}{2\omega}%
\operatorname{Im}\left\{  \overrightarrow{E}^{\ast}\cdot\left(  \nabla
\overrightarrow{E}\right)  \right\}  ,\quad\overrightarrow{P}_{se}=\frac{1}%
{2}\varepsilon_{0}\frac{c^{2}}{2\omega}\nabla\times\operatorname{Im}\left\{
\overrightarrow{E}^{\ast}\times\overrightarrow{E}\right\}  .
\end{split}
\end{align}
A "democratic" decomposition \cite{berry2009optical} involves both the electric and magnetic fields:%
\begin{align} \label{5}
\begin{split}
\overrightarrow{P}\left(  \overrightarrow{r}\right)    & =\overrightarrow{P}%
_{o}+\overrightarrow{P}_{s}\,; \\ 
\overrightarrow{P}_{o}\,  & =\frac{1}{2}\left(  \overrightarrow{P}%
_{oe}+\overrightarrow{P}_{om}\right)  ,\quad\overrightarrow{P}_{s}\,=\frac
{1}{2}\left(  \overrightarrow{P}_{se}+\overrightarrow{P}_{sm}\right)  \,;\\ 
\overrightarrow{P}_{om}  & =\mu_{0}\frac{c^{2}}{2\omega}\operatorname{Im}%
\left\{  \overrightarrow{H}^{\ast}\cdot\left(  \nabla\overrightarrow{H}%
\right)  \right\}  ,\quad\overrightarrow{P}_{sm}=\frac{1}{2}\mu_{0}\frac
{c^{2}}{2\omega}\nabla\times\operatorname{Im}\left\{  \overrightarrow{H}%
^{\ast}\times\overrightarrow{H}\right\}  .
\end{split}
\end{align}

In general, $\overrightarrow{P}_{oe}\neq\overrightarrow{P}_{om}$ and
$\overrightarrow{P}_{se}\neq\overrightarrow{P}_{sm}$ even in units where
$\varepsilon_{0}=\mu_{0}=c=1$ which are assumed throughout this paper.
 Note that in the case of TE field, which has only one nonzero component
of the electric field, say $E_{y}$, the electric part of the spin flow
$\overrightarrow{P}_{se}$ vanishes and in the case of an analogous TM field, the
magnetic part of the spin flow $\overrightarrow{P}_{sm}$ vanishes.

\subsection{Linear TE and TM polarization}
To construct the TE field from the scalar wavefunction Eq.~(\ref{Wanalyt}), we define the magnetic 
Hertz vector as $\overrightarrow{\Pi}_m=(0,0,1)W(x,z)$. Correspondingly, electric 
and magnetic fields are
\begin{equation} \label{TEem}
\overrightarrow{E}=i k c \mu_{0}\nabla\times\overrightarrow{\Pi}_m; \quad \quad
\overrightarrow{H}=\nabla\times\nabla\times\overrightarrow{\Pi}_m.
\end{equation}

With these fields the spin component $\overrightarrow{P}_{s}$ of the Poyinting vector and its electric
and magnetic parts, respectively, $\overrightarrow{P}_{se}$ and  $\overrightarrow{P}_{sm}$ were computed
according to Eq.~(\ref{5}). 
As noted above and like in \cite{kotlyar2026reverse}, $\overrightarrow{P}_{se}=0$, but the spin component
as a whole does not vanish because $\overrightarrow{P}_{sm}\neq0$

\begin{figure}[h]
\centering
\includegraphics[width=14cm]{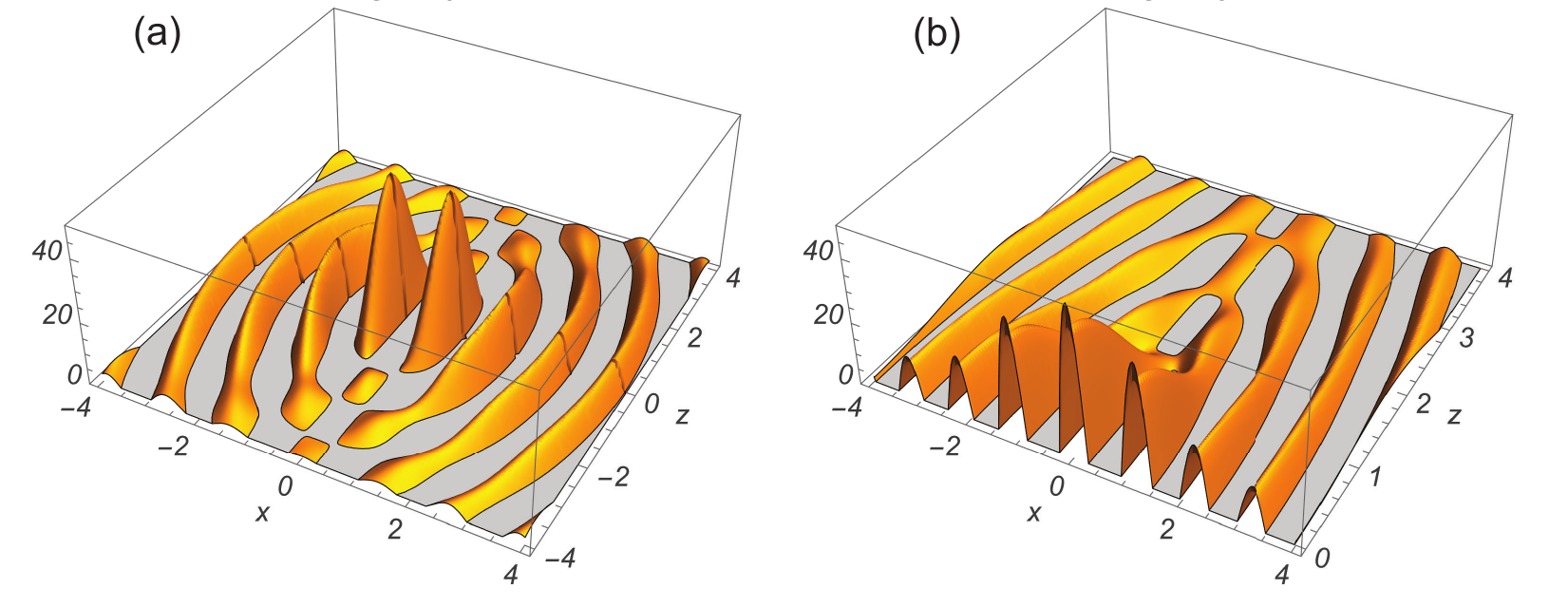}\caption{Plot of the longitudinal ($z$) projection of the magnetic 
part of the spin component $\overrightarrow{P}_{sm}$ of the Poynting vector, see Eq. (\ref{5}); (a), only 
positive values of the projection are shown, (b), only negative values, i.e., the backflow, are shown. 
The unit of spatial coordinate scales is $\lambda/2$. The vertical scale is relative and depends on 
electromagnetic units.} \label{Psmz}
\end{figure}

\begin{figure}[h!]
\centering
\includegraphics[width=15cm]{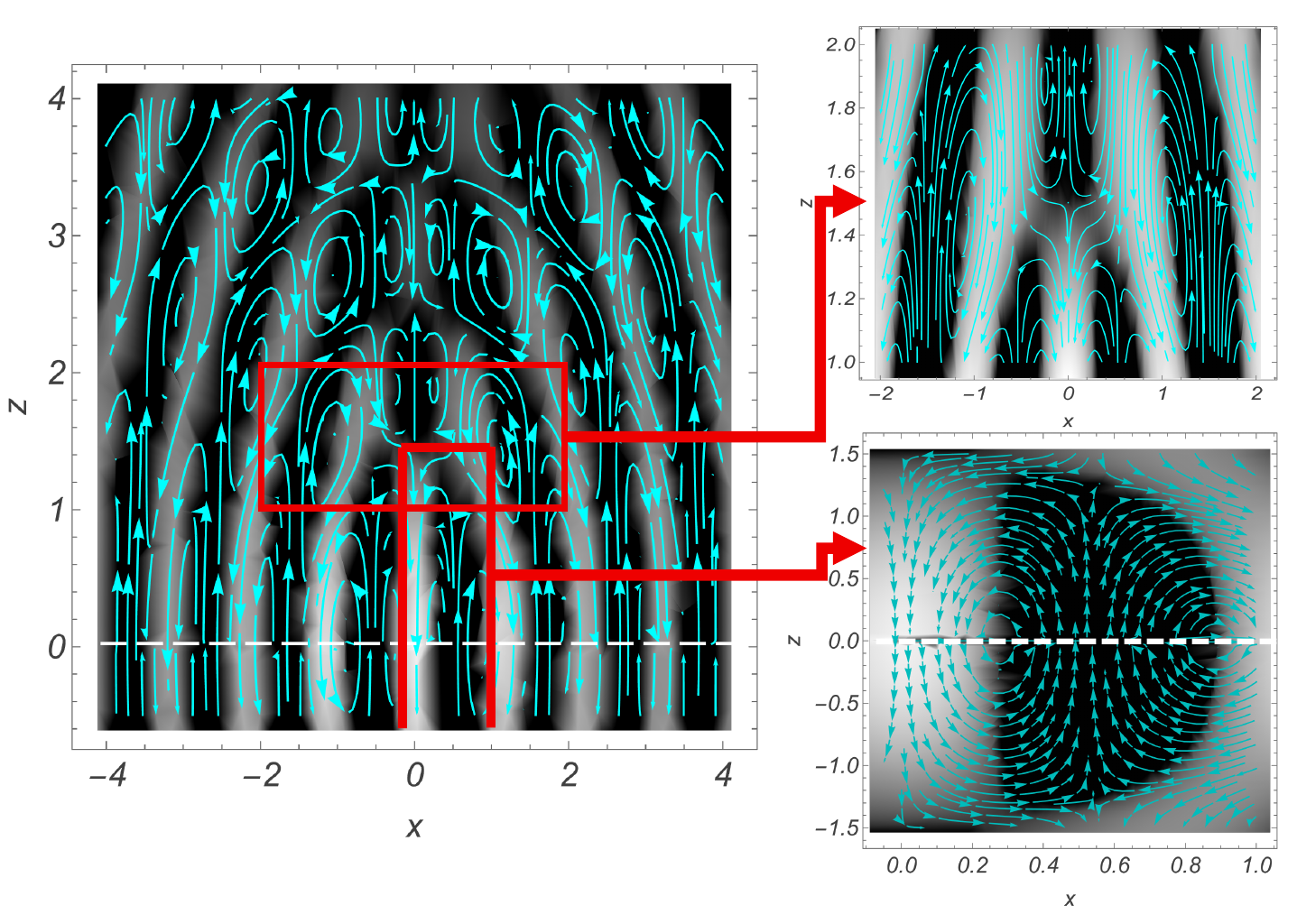}\caption{Greyscale density plot of negative values of the 
longitudinal  $z$-projection (backflow) $(-P_{sz})^{\frac{1}{3}}$ of the spin component of the 
Poynting vector, superimposed by a stream plot of the 
vector field $(P_{sx},P_{sz})$. Insets on the right have been plotted with higher resolution and without
preserving their aspect ratios. The greyscale plot presents the same quantity as in figure \ref{Psmz}b
since the electric part of the spin flow is absent and therefore the spin flow $\overrightarrow{P}_{s}$ consists solely of the magnetic part $\overrightarrow{P}_{sm}$.} \label{4spin}
\end{figure}

Figure \ref{Psmz} demonstrates a surprising result: the spin backflow is as strong as the forward flow. Commonly, the 
energy backflow in light fields, if it exists at all, is by an order or more weaker than the forward flow.
Parenthetically, we use the abbreviation 'spin backflow' because physically spin does not transport energy.
The backflow is the strongest near the origin where the forward flow vanishes. The reason of the latter is 
that although $|W(x,z)|^2$ has its maximum at the origin, the EM fields are expressed as spatial derivatives
of $W(x,z)$. Another peculiarity of the spin backflow is that unlike the scalar current backflow---
and below we see that also unlike the orbital backflow---it occurs along river-like areas instead of 
isolated island-like locations, see figure \ref{4spin}.

Next we consider the orbital energy flux (canonical energy flow). To save space, we omit here plots 
of positive values of the $z$-projections of $\overrightarrow{P}_{oe} $ and $\overrightarrow{P}_{om} $ 
defined in Eq.~(\ref{5}). Along the line $z=0$ the latter resembles figure \ref{Psmz}b while the former 
is zero at the origin followed by a double maximum along the line $z=0$. Generally, where the former has 
a crest of a wave, the latter almost vanishes, and \textit{vice versa}. 
This is because electric and magnetic fields are defined in Eq.~(\ref{TEem}) \textit{via} different orders
of spatial derivatives of the wavefunction. In terms of strength, both parts of the forward flow are 
more or less equal with each other, and also are equal to the strength of spin flow in figure \ref{Psmz}a.

\begin{figure}[h]
\centering
\includegraphics[width=14cm]{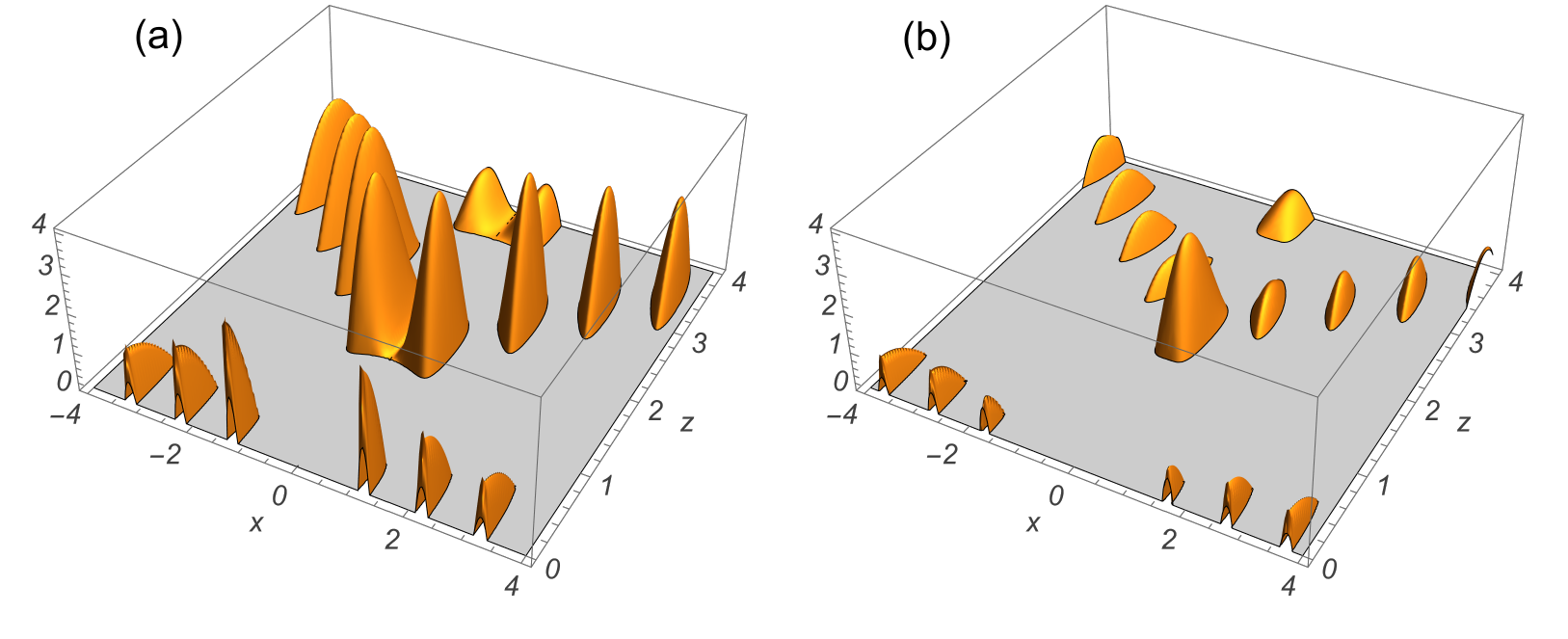}\caption{Plot of negative values of the longitudinal 
($z$)-projections (backflow) of (a) $\overrightarrow{P}_{oe} $ and (b)$\overrightarrow{P}_{om} $.}
\label{minusPoemz}%
\end{figure}

Figure \ref{minusPoemz} shows the spatial distribution of negative values of the $z-$projections of electric and 
magnetic parts of orbital energy backflow. They are by an order of magnitude weaker than those of 
the forward flow. Again, we see mutually exclusive locations of the backflow maxima and minima 
in the plots of electric part and magnetic part.

\begin{figure}[h]
\centering
\includegraphics[width=14cm]{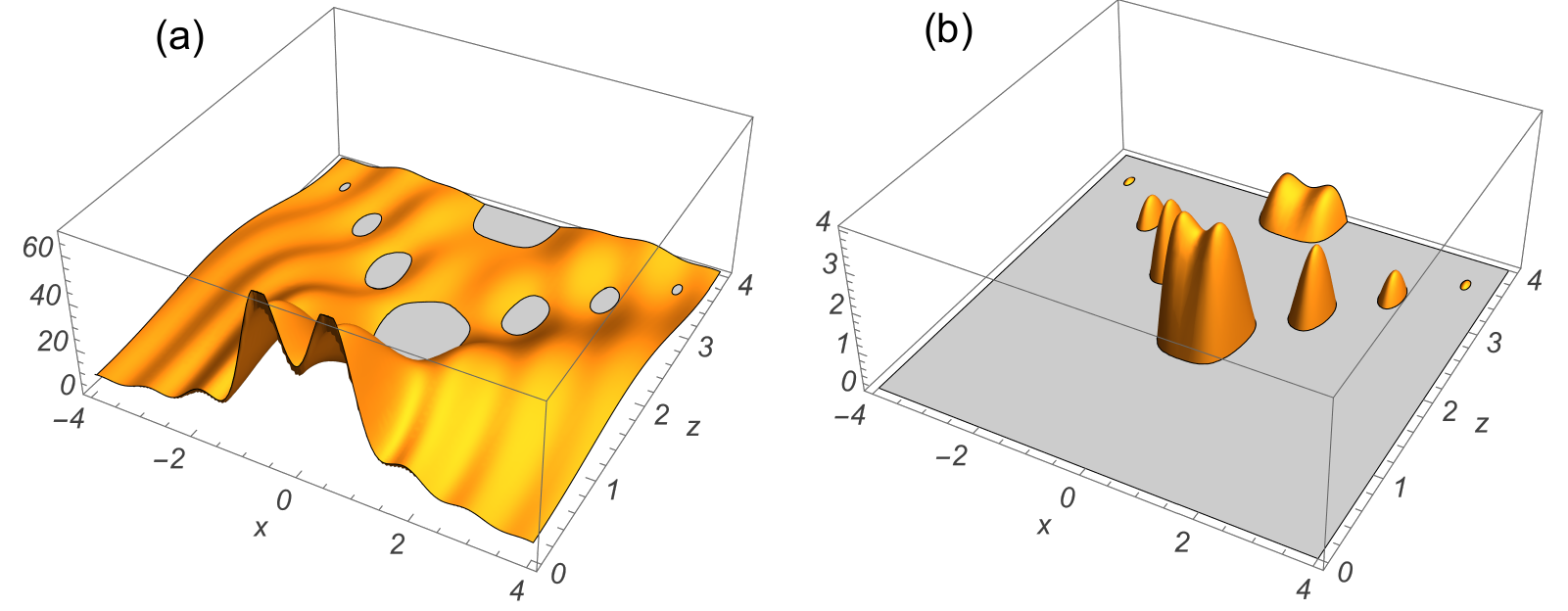}\caption{Plot of the longitudinal ($z$)-projection  of 
orbital contribution $\overrightarrow{P}_{o}$ to the Poynting vector, see Eq.~(\ref{5});~
(a), only positive values of the projection are shown, (b), only negative values, i.e., the backflow 
shown.} \label{Poz}
\end{figure}

Finally, figure \ref{Poz} shows the full orbital contribution to forward and backward flow density according
to averaging $\overrightarrow{P}_{o}= \left(\overrightarrow{P}_{oe}+\overrightarrow{P}_{om} \right)/2$ 
over the electric and magnetic parts. 
It follows from figure \ref{Poz} that despite the mutually alternating location of maxima and minima along 
the line $z=0$, the total forward flow
of $\overrightarrow{P}_{oe}$ and $\overrightarrow{P}_{om}$ is sufficiently strong in the region $0<z<1$,
thus suppressing the backflow.

Finally, there is no need to present here the plot of the negative values of the 
$z$-projection $P_z$ of the total Poynting 
vector $\overrightarrow{P}$ because it is exactly the same with $P_{oez}$ shown in 
figure \ref{minusPoemz}a, the latter arising from the equality $\overrightarrow{P}_{se}=0$. It follows
from Eqs.~(\ref{3})-(\ref{5}) that, alternatively, 
$P_{z}=(P_{oez}+P_{omz}+P_{smz})/2$ based on the electric-magnetic democracy discussed earlier.

\begin{figure}[h!]
\centering
\includegraphics[width=15cm]{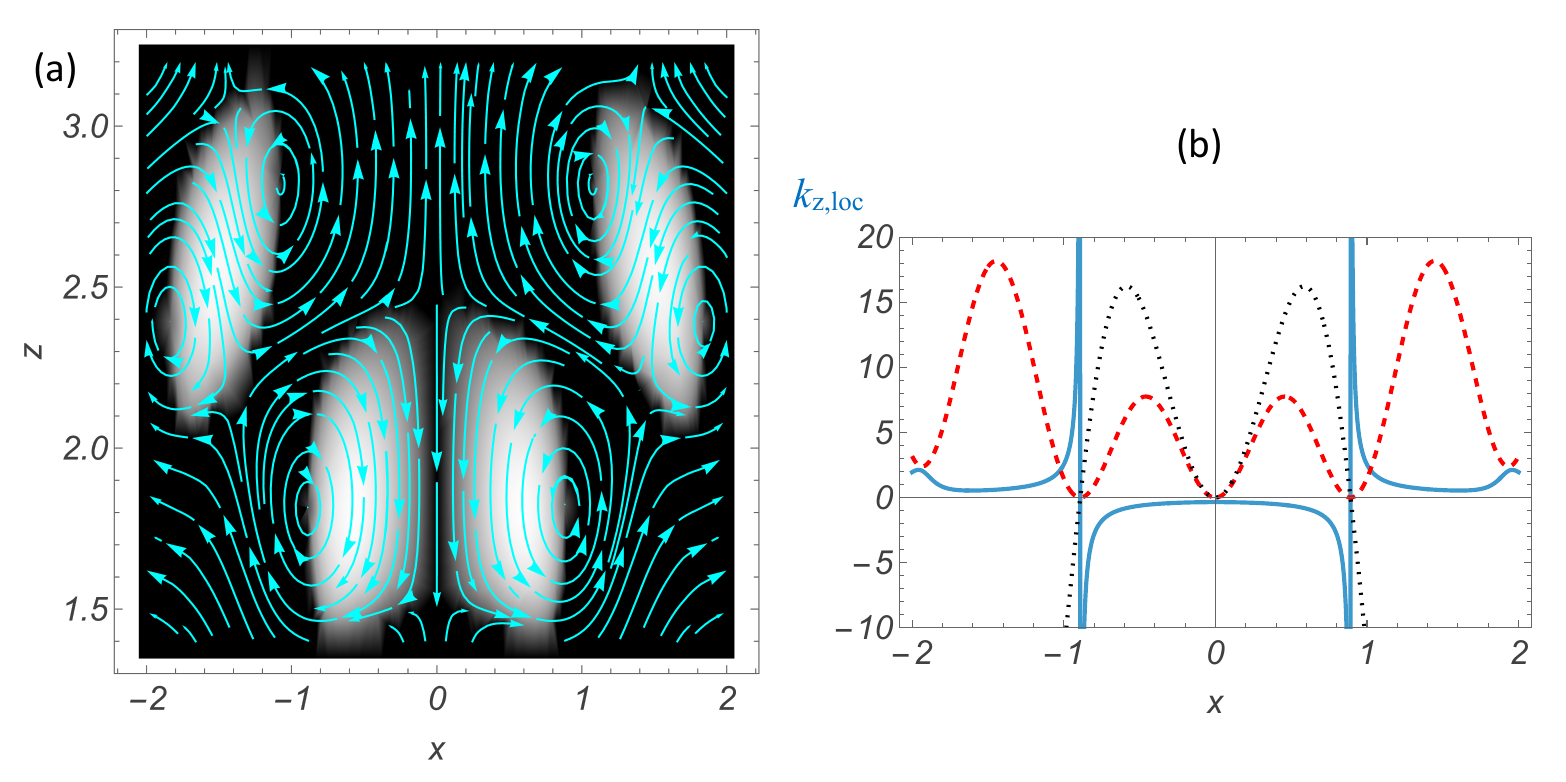}\caption{(a), greyscale density plot of negative values of the 
longitudinal  $z$-projection (backflow) $(-P_{z})^{\frac{1}{3}}$ of the 
Poynting vector, superimposed by a stream plot of the vector field $(P_{x},P_{z})$;
(b), dependencies of: $ k_{z,loc}\equiv\partial\Phi_{Ey}(x, z)/\partial z$
(solid curve) in comparison with the electric intensity $|E_y|^2/5$ (dashed curve), and with negative 
values of the longitudinal  $z$-projection (backflow) $4\times(-P_{z})$ (dotted curve), respectively; all 
dependencies are along the line $z=1.8$ which passes through the centers of the lower pair of vortices.
} \label{P-TEM}
\end{figure}

Figure \ref{P-TEM} shows the flow of the Poynting vector $\overrightarrow{P}$ in the central region 
of figure \ref{minusPoemz}a. As was the case with the scalar field and the spin flow, the vortices 
appear precisely on the borderline between forward and backward flows. Remarkable is that the 
halves of the backflow doublet are so close that the forward flow is pushed out from the slit 
between them as if the doublet is a sort of obstacle for the forward flow.

In distinction from the complex-valued scalar field whose phase was the only quantity for expressing the 
local wavenumber, here we use the phase $\Phi_{Ey}(x, z)=-i\ln(E_y/|E_y|)$ of the complex-valued electric
field for correlating with vortices in the Poynting vector flow.  We see in figure \ref{P-TEM}b that 
locations of vortices precisely coincide with joint points of singularities of 
$\partial\Phi_{Ey}(x, z)/\partial z$, flips of the sign of $P_z$, and zeros of $|E_y|^2$.  
The polarity of $k_{z,loc}$ at all values of $x$ is the same as that of $P_z$. In other words, directions
of local wave vector and energy flow coincide as they have to.
Notice also that there are no vortices or large values of the local wavenumber
between the halves of the doublet, despite $|E_y|^2$ being exactly equal 
to zero at the point $(0,1.8)$.

Additionally, we carried out the same study on the TM field, which was constructed using the electric 
Hertz vector $\overrightarrow{\Pi}_e=(0,0,1)W(x,z)$. The results turned out to be the same
if one interchanges the indices "e" and "m" in all pertinent quantities and plots. This is a manifestation
of the known symmetry between electric and magnetic fields in free space.

\subsection{Circular polarization}
We constructed the circularly polarized field using the magnetic Hertz vector $\overrightarrow{\Pi}_m=(1,i,0)W(x,z)$. Generally, the characteristics
of the backflow are not much different from the cases of TE and TM polarization.

\begin{figure}[h!]
\centering
\includegraphics[width=15cm]{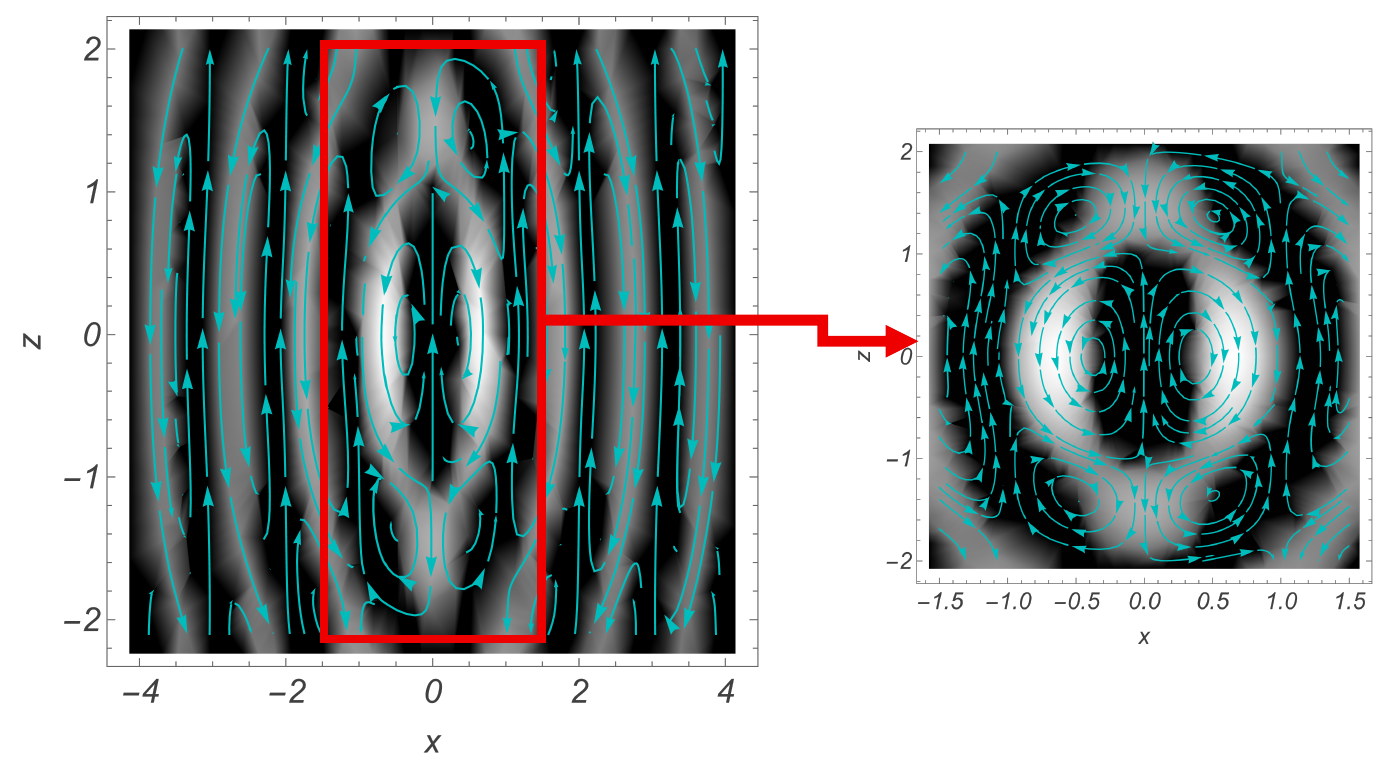}\caption{Greyscale density plot of negative values of the 
longitudinal  $z$-projection (backflow) $(-P_{sz})^{\frac{1}{3}}$ of the spin component of the 
Poynting vector of the circularly polarized field, superimposed by a stream plot of the 
vector field $(P_{sx},P_{sz})$. Inset on the right has been plotted with higher resolution and with
an aspect ratio changed toward equal coordinate scales.} \label{cicspin}
\end{figure}

The electric part of the spin backflow is only two times weaker than the electric part of the spin 
forward flow. The same ratio holds for the magnetic part, which, however, is 5 times weaker
than the electric part when comparing their maxima at the origin. Hence, both parts together result in
backflow which is nearly half of the forward flow. Let us recall that in the TE field the spin backflow
is as strong as the forward flow. An explanation of this difference is that in the TE field the electric
part of spin flow is absent. Another difference is that, as seen in figure \ref{cicspin}, the backflow
exactly at the origin is absent while in the TE field it is maximum there, see figure \ref{Psmz}b.
It is also very interesting to note that according to figure \ref{cicspin} the spin flow cannot leave 
the region of origin, remaining to circulate there.

\begin{figure}[h]
\centering
\includegraphics[width=14cm]{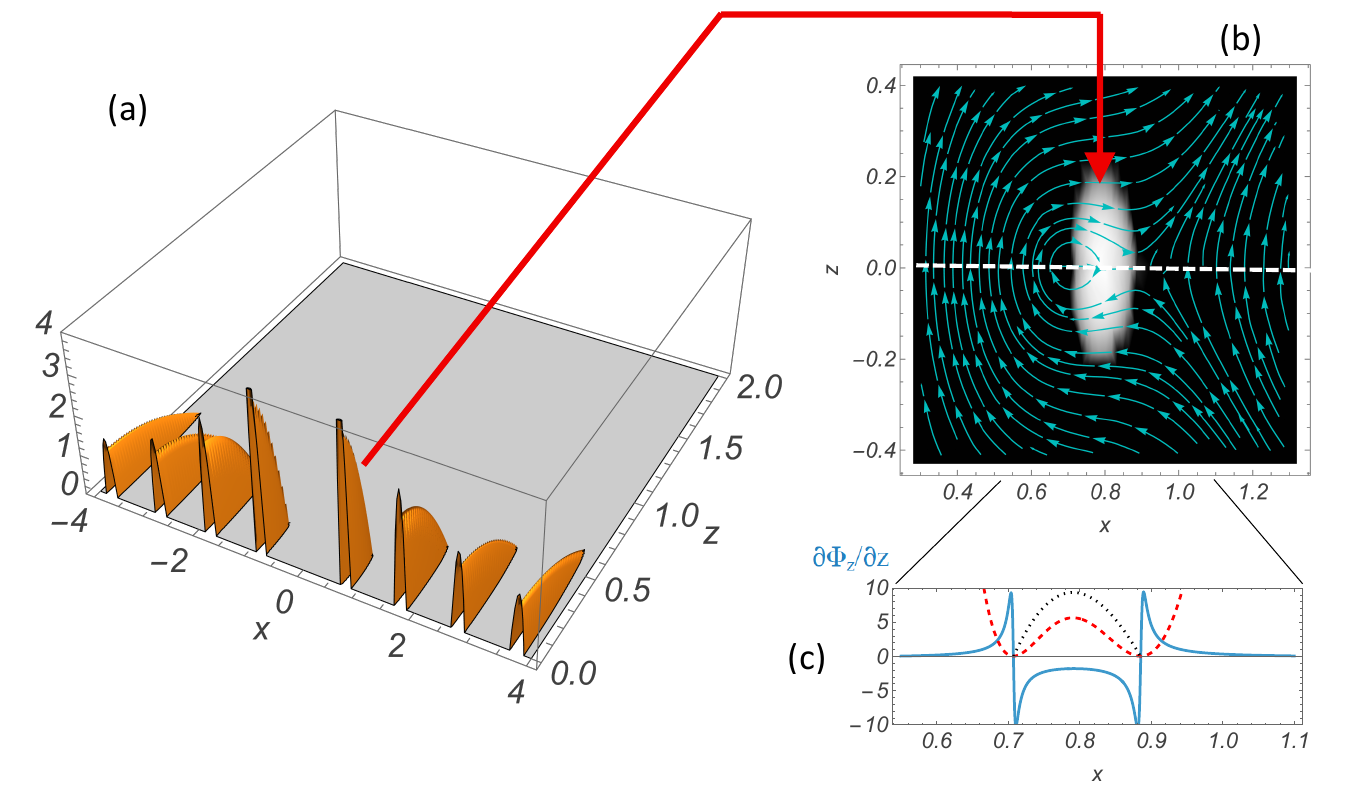}\caption{(a), plot of negative values of the longitudinal 
($z$)-projection (backflow) of  the Poynting vector $P_{z}$ of the circularly 
polarized field; (b), greyscale density plot of one peak in it, superimposed by a stream plot of the 
vector field $(P_{x},P_{z})$; (c), dependencies of: $\partial\Phi_z(x, z)/\partial z\,/5$
(solid curve) in comparison with the intensity $|Q_z(x, z)|^2\,/10$ (dashed curve), and with negative 
values of the longitudinal  $z$-projection (backflow) $2.5\times(-P_{z})$ (dotted curve), respectively; all 
dependencies are along the line $z=0.01$; peak values of $\partial\Phi_z(x, z)/\partial z$ reach $\pm 50$.}
\label{circP}%
\end{figure}

Finally, figure \ref{circP} shows the backflow of the total Poynting vector. Positive values of 
both the electric and magnetic parts of the orbital components are strong exceeding negative values about
two orders of magnitude (peak of their sum at the origin reaches 300 units). Therefore, the only 
contribution  to the total backflow, seen in figure \ref{circP} comes from peaks of the spin 
component along the line $z=0$. As can be seen from the vertical scale of the 3D plot and from 
comparison of the insets in figures \ref{cicspin} and \ref{circP}, the backflow peaks are suppressed 
about ten times and reshaped by the orbital forward flow. 

In contrast to the scalar field and TE field, in the given case there is no quantity at hand 
for expressing the local wavenumber because the fields have more than one nonzero component. 
Therefore, we define a complex-valued Poynting vector as 
$\overrightarrow{Q}=\overrightarrow{E}^*\times \overrightarrow{H}$
and the phase $\Phi_z(x, z)$ of its projection $Q_z$ to the axis $z$  as $\Phi_z(x, z)=-i 
\ln(Q_z/|Q_z|)$. We see in figure \ref{P-TEM}b that locations of vortices precisely coincide 
with joint points of singularities of $\partial\Phi_z(x, z)/\partial z$, 
flips of the sign of $P_z$, and zeros of $|Q_z|^2$. However, unlike the case of scalar fields, there
is no simple relation between the local wavevector and the Poynting vector, as in Eq.~(\ref{J2x}). 
Instead, we have the relation $P_z=\Re(Q_z)/2=|Q_z|\cos \Phi_z/2$ from which it follows that $P_z$
and $\partial\Phi_z(x, z)/\partial z$ are not necessarily of the same sign. Nevertheless we observe in
figure \ref{P-TEM}b that the direction of $\partial\Phi_z(x, z)/\partial z$ is opposite to backflow.
The curves in figure \ref{circP}c again 
demonstrate that the locations of singularly large local wavenumbers, intensity zeros, and forward-backward 
flow boundaries coincide on the line $z=0.01$ (exactly $z=0$ was not chosen due to computational problems
that arise if both wavefunctions in Eq.~(\ref{Wanalyt}) and Eq.~(\ref{z=0}) are involved in the 
calculations). But there
is a vortex at the left-hand-side joint point. At the right-hand-side joint point there is, as seen in
figure \ref{circP}, a saddle point. This observation was confirmed by making a high-resolution stream plot
of an extra small sub-wavelength region $(0.85<x<0.95, -0.06<z<0.06)$. It is known from fluid dynamics 
that in addition to vortices a flow can contain hyperbolic points (saddle or stagnation points) where 
also the intensity  vanishes and local wavenumbers diverge. Looking more closely at the stream plots 
in the preceding sections we see that saddle points alternate with vortices in every field, as it should be. 

To summarize, it must be concluded, however, that if both electric and magnetic parts of the orbital 
and spin components are present, they trend to mutually suppress maxima and minima of each other. As a 
result, while circular polarization in some types of waves fosters the backflow, 
this is not true in the given case.

\section{Concluding remarks}

Energy backflow is a counterintutive effect in the physics of homogeneous waves propagating in free
space without singularities and is essential for various physical phenomena and applications in which 
the direction of the Poynting vector is important.  Especially intriguing is the effect in the so-called
unidirectional wave fields all plane-wave constituents of which are directed forward. Therefore, 
of particular recent interest has been  the energy backflow  in pulsed and monochromatic 
unidirectional electromagnetic waves.

Our work in this article has been focused on the orbital and spin current density backflow  in 
the case of an analytical unidirectional monochromatic electromagnetic field in vacuum. We were 
motivated in this direction by the work by Kotlyar \textit{et al.}\cite{kotlyar2026reverse,kotlyar20262d}
who studied  orbital reverse flows in 2D nonparaxial monochromatic unidirectional TE fields. 
We briefly summarize the further developments in our study compared to their work for clarity.	

Our wave function is an exact analytical expression not only on the $x$ axis as in Refs.
\cite{kotlyar2026reverse,kotlyar20262d} but over 
the whole 2D space. This provided us great facility in computing the Poynting vector, as well as 
all its electric and magnetic constituents, and graphing 3D and streamline plots for all relevant 
values of propagation distance $z$ without resorting to time-consuming numerical integrations. 

By using the Hertz vector formalism, we constructed consistently and studied TE, TM, and circularly 
polarized EM fields instead of limiting the calculations to a TE  field with a single component of 
the electric field along the y-direction. If this electric field, chosen  in an \textit{ad hoc manner}
as in Refs. \cite{kotlyar2026reverse,kotlyar20262d},
is used to determine the monochromatic magnetic field using the relationship 
$\overrightarrow H=\nabla \times \overrightarrow E_m/(ikc \mu_0)$ 
from Maxwell’s first equation, the remaining Maxwell equations are not necessarily satisfied. 

An important contribution in our article is the use of an “electromagnetic democracy,” whereby 
the spin and orbital current density components of the Poynting vector are expressed in terms 
of both electric and magnetic fields.

The main results in the article can be summarized as follows. In the case of a scalar wave,  
vortices with very large values of the local wavenumber appear exactly on the borderline between 
forward and backward flows, specifically at points where the modulus of the wavefunction becomes zero, 
thus demonstrating explicitly known phenomena of singular optics. While commonly energy backflow in 
scalar and vector-valued electromagnetic fields is more than an order of magnitude weaker than 
forward flow, for our  2D unidirectional field the spin backward flow is as strong as the forward flow.
Also, the magnetic and electric parts of the orbital and spin flow in TE and TM fields, respectively,  
exhibit backflows of comparable strengths.

Physically important  is the presence of backflow in the total Poynting vector, and not necessarily 
in the backflow behavior of its spin and orbital current density components. Although the  latter 
exhibit forward and backward flows of the same order of magnitude, the “composite” energy backflow 
in the Poynting vector is very small and distributed differently in the case of TE/TM and circular
polarizations. In the former it appears on  the entire $x-z$ plane, with diminishing  strength for  
increasing $x, y$ values away from the origin. For circular polarization  the energy backflow is 
restricted only in a region of small values of $z$.

Another important contribution in this article is the study of the correlation of the position of 
energy backflows with local wavenumbers and their signs, zeros of appropriate intensities, and the 
presence of vortices.

We believe that our results not only contribute to the understanding  of phenomena of singular optics
but also could be useful for applications such as microparticle manipulation when placed in a light 
beam, as well as in nonparaxial generalizations of reverse energy behavior near the focus of a beam.

\bibliographystyle{unsrt}
\addcontentsline{toc}{section}{\refname}
\bibliography{Bibliography0.bib}

@article{ignatovsky1919diffraction,
  title={Diffraction by a lens having arbitrary opening},
  author={Ignatovsky, VS},
  journal={Trans. Opt. Inst. Petrograd},
  volume={1},
  number={4},
  year={1919}
}

@article{richards1959electromagnetic,
  title={Electromagnetic diffraction in optical systems, II. Structure of the image field in an aplanatic system},
  author={Richards, Bernard and Wolf, Emil},
  journal={Proceedings of the Royal Society of London. Series A. Mathematical and Physical Sciences},
  volume={253},
  number={1274},
  pages={358--379},
  year={1959},
  publisher={The Royal Society London}
}

@article{katsenelenbaum1997direction,
  title={What is the direction of the {P}oynting vector?(A methodic note)},
  author={Katsenelenbaum, BZ},
  journal={Journal of communications technology \& electronics},
  volume={42},
  number={2},
  pages={119--120},
  year={1997}
}

@article{novitsky2007negative,
  title={Negative propagation of vector Bessel beams},
  author={Novitsky, Andrey V and Novitsky, Denis V},
  journal={Journal of the Optical Society of America A},
  volume={24},
  number={9},
  pages={2844--2849},
  year={2007},
  publisher={Optical Society of America}
}

@article{salem2011energy,
  title={Energy flow characteristics of vector X-waves},
  author={Salem, Mohamed A and Ba{\u{g}}c{\i}, Hakan},
  journal={Optics express},
  volume={19},
  number={9},
  pages={8526--8532},
  year={2011},
  publisher={Optical Society of America}
}

@article{vaveliuk2012negative,
  title={Negative propagation effect in nonparaxial Airy beams},
  author={Vaveliuk, Pablo and Martinez-Matos, Oscar},
  journal={Optics Express},
  volume={20},
  number={24},
  pages={26913--26921},
  year={2012},
  publisher={Optical Society of America}
}

@article{mitri2016reverse,
  title={Reverse propagation and negative angular momentum density flux of an optical nondiffracting nonparaxial fractional Bessel vortex beam of progressive waves},
  author={Mitri, FG},
  journal={Journal of the Optical Society of America A},
  volume={33},
  number={9},
  pages={1661--1667},
  year={2016},
  publisher={Optical Society of America}
}

@article{yuan2019plasmonics,
  title={“Plasmonics” in free space: observation of giant wavevectors, vortices, and energy backflow in superoscillatory optical fields},
  author={Yuan, Guanghui and Rogers, Edward TF and Zheludev, Nikolay I},
  journal={Light: Science \& Applications},
  volume={8},
  number={1},
  pages={2},
  year={2019},
  publisher={Nature Publishing Group UK London}
}

@article{kotlyar2019energy,
  title={Energy backflow in the focus of a light beam with phase or polarization singularity},
  author={Kotlyar, VV and Stafeev, SS and Nalimov, AG},
  journal={Physical Review A},
  volume={99},
  number={3},
  pages={033840},
  year={2019},
  publisher={APS}
}

@misc{khonina2022tailoring,
  title={Tailoring of Inverse Energy Flow Profiles with Vector Lissajous Beams. Photonics 2022, 9, 121},
  author={Khonina, SN and Porfirev, AP and Ustinov, AV and Kirilenko, MS and Kazanskiy, NL},
  year={2022},
  publisher={s Note: MDPI stays neu-tral with regard to jurisdictional claims in~…}
}

@article{kotlyar2020mechanism,
  title={Mechanism of formation of an inverse energy flow in a sharp focus},
  author={Kotlyar, VV and Stafeev, SS and Nalimov, AG and Kovalev, AA and Porfirev, AP},
  journal={Physical Review A},
  volume={101},
  number={3},
  pages={033811},
  year={2020},
  publisher={APS}
}

@article{li2020controlled,
  title={Controlled negative energy flow in the focus of a radial polarized optical beam},
  author={Li, Hehe and Wang, Chen and Tang, Miaomiao and Li, Xinzhong},
  journal={Optics Express},
  volume={28},
  number={13},
  pages={18607--18615},
  year={2020},
  publisher={Optical Society of America}
}

@article{saari2021backward,
  title={Backward energy flow in simple four-wave electromagnetic fields},
  author={Saari, Peeter and Besieris, Ioannis},
  journal={European Journal of Physics},
  volume={42},
  number={5},
  pages={055301},
  year={2021},
  publisher={IOP Publishing}
}

@article{geints2022simulation,
  title={Simulation of enhanced optical trapping in a perforated dielectric microsphere amplified by resonant energy backflow},
  author={Geints, Yury E and Minin, Igor V and Minin, Oleg V},
  journal={Optics Communications},
  volume={524},
  pages={128779},
  year={2022},
  publisher={Elsevier}
}

@article{bialynicki2022backflow,
  title={Backflow in relativistic wave equations},
  author={Bialynicki-Birula, Iwo and Bialynicka-Birula, Zofia and Augustynowicz, Szymon},
  journal={Journal of Physics A: Mathematical and Theoretical},
  volume={55},
  number={25},
  pages={255702},
  year={2022},
  publisher={IOP Publishing}
}

@article{besieris2023energy,
  title={Energy backflow in unidirectional spatiotemporally localized wave packets},
  author={Besieris, Ioannis and Saari, Peeter},
  journal={Physical Review A},
  volume={107},
  number={3},
  pages={033502},
  year={2023},
  publisher={APS}
}

@article{ghosh2024canonical,
  title={Canonical and {P}oynting currents in propagation and diffraction of structured light: tutorial},
  author={Ghosh, Bohnishikha and Daniel, Anat and Gorzkowski, Bernard and Bekshaev, Aleksandr Y and Lapkiewicz, Radek and Bliokh, Konstantin Y},
  journal={Journal of the Optical Society of America B},
  volume={41},
  number={6},
  pages={1276--1289},
  year={2024},
  publisher={Optica Publishing Group}
}

@article{saari2024energy,
  title={Energy backflow in unidirectional monochromatic and space--time waves},
  author={Saari, Peeter and Besieris, Ioannis M},
  journal={Photonics},
  volume={11},
  number={12},
  pages={1129},
  year={2024},
  publisher={MDPI}
}

@article{ustinov2024interference,
  title={Interference generation of a reverse energy flow with varying orbital and spin angular momentum density},
  author={Ustinov, Andrey V and Porfirev, Alexey P and Khonina, Svetlana N},
  journal={Photonics},
  volume={11},
  number={10},
  pages={962},
  year={2024},
  publisher={MDPI}
}

@article{han2024controllable,
  title={Controllable reverse energy flow in the focus of tightly focused hybrid vector beams},
  author={Han, Lei and Qi, Jiale and Gao, Chuchu and Li, Fuli},
  journal={Optics Express},
  volume={32},
  number={21},
  pages={36865--36874},
  year={2024},
  publisher={Optica Publishing Group}
}

@article{Chun-Fang2026,
  title={Dependence of {P}oynting vector on state of polarization},
  author={You, Xiao-Lu and Li, Chun-Fang},
   journal={Photonics},
  volume={13},
  number={2},
  pages={137},
  year={2024},
  publisher={MDPI}
 }

@article{bekshaev2007transverse,
  title={Transverse energy flows in vectorial fields of paraxial beams with singularities},
  author={Bekshaev, A Ya and Soskin, MS},
  journal={Optics communications},
  volume={271},
  number={2},
  pages={332--348},
  year={2007},
  publisher={Elsevier}
}

@article{berry2009optical,
  title={Optical currents},
  author={Berry, Michael V},
  journal={Journal of Optics A: Pure and Applied Optics},
  volume={11},
  number={9},
  pages={094001},
  year={2009}
}

@article{bliokh2014extraordinary,
  title={Extraordinary momentum and spin in evanescent waves. Suppl. Mat.},
  author={Bliokh, Konstantin Y and Bekshaev, Aleksandr Y and Nori, Franco},
  journal={Nature communications},
  volume={5},
  number={1},
  pages={3300},
  year={2014},
  publisher={Nature Publishing Group UK London}
}

@article{kotlyar2025canonical,
  title={Canonical energy backflow in Airy beams},
  author={Kotlyar, Victor and Kovalev, Alexey and Nalimov, Anton},
  journal={Journal of the Optical Society of America B},
  volume={43},
  number={1},
  pages={128--134},
  year={2025},
  publisher={Optica Publishing Group}
}

@article{kotlyar20262d,
  title={2D optical vortices and a reverse energy flow occuring near the intensity zeros},
  author={Kotlyar, Victor and Nalimov, Anton and Kovalev, Alexey and Telegin, Alexey},
  journal={Optics Letters},
  volume={51},
  number={4},
  pages={973--976},
  year={2026},
  publisher={Optica Publishing Group}
}

@article{kotlyar2026reverse,
  title={Reverse canonical energy flow in the near field of a non-paraxial sinc beam without evanescent waves},
  author={Kotlyar, VV and Nalimov, AG and Kovalev, AA and Arfan, Muhammad and Stafeev, SS},
  journal={Journal of Optics},
  volume={28},
  number={1},
  pages={015605},
  year={2026},
  publisher={IOP Publishing}
}

@article{kotlyar2025backward,
  title={Backward canonical energy flow in the near field of three non-paraxial 2D beams},
  author={Kotlyar, VV and Kovalev, AA and Nalimov, AG and Telegin, AM},
  journal={Journal of the Optical Society of America A},
  volume={42},
  number={9},
  pages={1396--1402},
  year={2025},
  publisher={Optica Publishing Group}
}

@article{stafeev2026phase,
  title={Phase singularities near reverse energy flows in tightly focused optical vortices},
  author={Stafeev, Sergey S and Zaitcev, Vladislav D and Guo, K and Liu, B and Guo, Z and Kotlyar, Victor V},
  journal={Applied Optics},
  volume={65},
  number={4},
  pages={1042--1048},
  year={2026},
  publisher={Optica Publishing Group}
}

@article{kotlyar2026reverseAPB,
  title={Reverse canonical energy flux in a nonparaxial linearly polarized beam near the initial plane},
  author={Kotlyar, VV and Kovalev, AA and Nalimov, AG},
  journal={Applied Physics B},
  volume={132},
  number={1},
  pages={11},
  year={2026},
  publisher={Springer}
}

@misc{mandel1996optical,
  title={Optical coherence and quantum optics},
  author={Mandel, Leonard and Wolf, Emil and Shapiro, Jeffrey H},
  year={1996},
  publisher={American Institute of Physics}
}

\end{document}